\documentclass[preprints,article,accept,moreauthors,pdftex]{Definitions/mdpi} 

\firstpage{1} 
\pubvolume{1}
\issuenum{1}
\articlenumber{0}
\doinum{10.3390/a14030075}
\pubyear{2021}
\copyrightyear{2020}
\datereceived{22 January 2021} 
\dateaccepted{22 February 2021} 
\datepublished{26 February 2021} 
\hreflink{https://doi.org/10.3390/a14030075} 

\usepackage{booktabs}       
\usepackage{amsfonts}       
\usepackage{nicefrac}       
\usepackage{enumitem}	
\usepackage{float}
\usepackage{subfig}

\usepackage{framed} 

\graphicspath{
{../Figures/}
}

\DeclareUnicodeCharacter{2212}{-}

\DeclareMathOperator*{\argmin}{arg\,min}

\newcommand{\GRU}{\operatorname{GRU}}
\newcommand{\CE}{\operatorname{CrossEntropy}}
\newcommand{\X}{\mathbf{X}}
\def \x{\rm x}
\def \y{\rm y}
\def \z{\rm z}
\def \tp{\rm t}
\def \R{\mathbb{R}}
\usepackage{amsmath,amssymb}

\Title{A deep learning model for data-driven discovery of functional connectivity}

\Author{Usman Mahmood $^{1}$\orcidA{} *, Zening Fu $^{1}$, Vince D. Calhoun$^{1}$, Sergey Plis $^{1}$}

\AuthorNames{Usman Mahmood, Zening Fu, Vince D. Calhoun and Sergey Plis}

\address{%
$^{1}$ \quad Tri-institutional Center for Translational Research in Neuroimaging and Data Science (TReNDS), Georgia State University, Georgia Tech, Emory; umahmood1@student.gsu.edu, \{zfu,vcalhoun\}@gsu.edu,  s.m.plis@gmail.com\\}

\corres{Correspondence: umahmood1@student.gsu.edu}

\abstract{ Functional connectivity (FC) studies have demonstrated the overarching value of studying the brain and its disorders through the undirected weighted graph of fMRI correlation matrix. Most of the work with the FC, however, depends on the way the connectivity is computed, and further depends on the manual post-hoc analysis of the FC matrices. In this work we propose a deep learning architecture BrainGNN that learns the connectivity structure as part of learning to classify subjects. It simultaneously applies a graphical neural network to this learned graph and learns to select a sparse subset of brain regions important to the prediction task. We demonstrate the model's state-of-the-art classification performance on a schizophrenia fMRI dataset and demonstrate how introspection leads to disorder relevant findings. The graphs learned by the model exhibit strong class discrimination and the sparse subset of relevant regions are consistent with the schizophrenia literature.
}


\begin{document}


\section{Introduction}
Functional connectivity which is often computed using cross-correlation among brain regions of interest (ROIs) is a powerful approach which has been shown to be informative for classifying brain disorders and revealing putative bio-markers relevant to the underlying disorder~\cite{10.1093/brain/awn018, Lynall9477, https://doi.org/10.1002/hbm.23524, 10.1007/978-3-030-59728-3_52}.
Inferring and using functional connectivity through spatio-temporal data, e.g. functional magnetic resonance imaging (fMRI), has been an especially important area of research in recent times.
Functional connectivity can improve our understanding of brain dynamics and improve classification accuracy for brain disorders such as schizophrenia.
Recent work \citep{yan2017discriminating} uses functional network connectivity (FNC) as features to predict schizophrenia related changes.
Whereas, \citealt{PARISOT2018117} uses functional connectivity obtained by a fixed formula with phenotypic and imaging data as inputs and to extract graphic features for the classification of AD and Autism.
\citealt{BrainNetCNN} also uses connection strength between brain regions as edges, typically defined as the number of white-matter tracts connecting the regions. \citealt{10.1007/978-3-319-66182-7_54} employs spectral graph theory to learn similarity metrics among functional connectivity networks.

These papers, as well as many others, have shown the efficacy of functional
connectivity and feature extraction based on neural network models.
However, existing studies often heavily depend on the underlying
method of functional connectivity estimation, in terms of classification accuracy, feature extraction, or learning brain dynamics. Studies like \cite{RASHID2016645,Saha2020.06.24.161745,SALMAN2019101747} depend on hand-crafted features based on methods like ICA (Independent Component Analysis). These studies work very well on classification but do not learn a sparse graph and not helpful for identifying bio-markers in the brain. 

Many functional connectivity studies~\cite{10.3389/fnins.2018.00525} on brain disorders utilize ROIs predefined based on anatomical or functional atlases, which are either fixed for all subjects or based are based on group differences.

These approaches ignore the possibility of inter-subject variations of ROIs, especially the variations due to the underlying disease conditions.
They also rely on the complete set of these ROIs discounting the possibility that only a small subset may be important at a time.
A disorder can have varying symptoms for different people, hence making it crucial to determine disorder and subject specific ROIs. 

In this work, we address the problems of using a fixed method of learning functional connectivity and using it as a fixed graph to represent brain structure (the standard practices) by utilizing a novel attention based Graph Neural Network (GNN) ~\cite{li2016gated}, which we call BrainGNN. We apply it to fMRI data and 1) achieve comparable classification accuracy to existing algorithms, 2) learn dynamic graph functional connectivity, and 3) increase model interpretability by learning which regions from the set of ROIs are relevant for the classification, enabling additional insights into the health and disordered brain.


\section{Materials and Methods}

\subsection{Materials}

In this study, we worked with the data from Function Biomedical Informatics Research Network (FBIRN)~\cite{keator2016function} dataset including schizophrenia (SZ) patients and healthy controls (HC) for testing our model. Details of the dataset are in the following section.

\subsubsection{FBIRN}
Resting fMRI data from the phase III FBIRN were analyzed for this project. The dataset has $368$ total subjects out of which $311$ were selected based on the preprocessing method explained in \ref{preprocessing}.

\subsubsection{Preprocessing}
\label{preprocessing}
The fMRI data was preprocessed using statistical parametric mapping (SPM12, \url{http://www.fil.ion.ucl.ac.uk/spm/}) under the MATLAB 2019 environment. A rigid body motion correction was performed to correct subject head motion, followed by the slice-timing correction to account for timing difference in slice acquisition. The fMRI data were subsequently warped into the standard Montreal Neurological Institute (MNI) space using an echo planar imaging (EPI) template and were slightly resampled to $3 \times 3 \times 3$ mm$^3$ isotropic voxels. The resampled fMRI images were then smoothed using a Gaussian kernel with a full width at half maximum (FWHM) = $6$ mm. After the smoothing, the functional images were temporally filtered by a finite impulse response (FIR) bandpass filter (0.01 Hz-0.15 Hz). Then for each voxel, six rigid body head motion parameters, white matter (WM) signals, and cerebrospinal fluid (CSF) signals were regressed out using linear regression.

We selected subjects for further analysis \cite{FU2021117385} if the subjects have head motion $\le 3^\circ$ and $\le 3$ mm, and with functional data providing near full brain successful normalization~\cite{fu2019altered}.

This resulted in a total of $311$ subjects with $151$ healthy controls and $160$ subjects with schizophrenia. Each subject is represented by $\X \in \R^{\x \times \y \times \z \times \tp}$, where $\x, \y, \z$ represent the number of voxels in each dimension and $\tp$ is the number of time points which are $160$. To reduce the affect of noise we zscore the time sequence of each voxel independently. Thus, time series of every voxel is replaced by the z-score of the time series. This does not have any affect on the data dimensions. 

To partition the data into regions use automated anatomical labeling (AAL) \cite{TZOURIOMAZOYER2002273} which contains $116$ brain regions. Taking sum of the voxels inside a region is an easy and common method but this gives and unfair advantage to bigger regions. For this, we take the weighted average of the voxel intensities inside a region. Weight is the value of a voxel being inside a region, as these values are not binary. Averaging helps to negate the bias towards bigger regions. This results in a dataset $D = {(S_1, S_2, S_3 ...... S_n)}$ where $S_i \in \R^{r \times \tp}$, $n=311$, $r=116$, $t=160$.

\subsection{Method}

\begin{figure}[H]
\centering
\includegraphics[width=\linewidth]{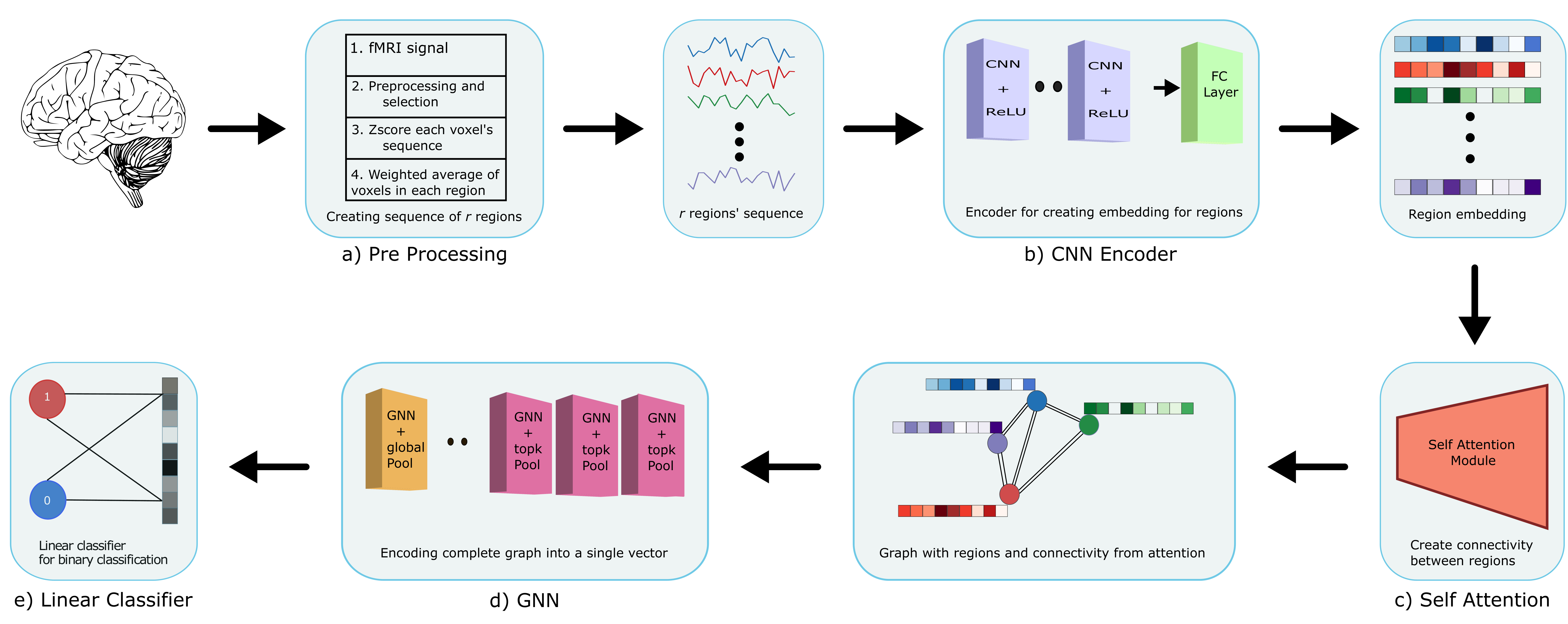}
\caption{BrainGNN architecture using a) Preprocessing: To preprocess the raw data with different steps (\ref{preprocessing}). b) 1DCNN: To create embedding for regions (\ref{Encoder}). c) Self-attention: To create connectivity between regions (\ref{Self-Attention}) d) GNN: To obtain a single feature vector for the entire graph (\ref{GNN}) and e) Linear classifier: To obtain the final classification.}
\label{fig:BrainGNNArch}
\end{figure}

We have three distinct parts in our novel attention based GNN architecture: 1) a Convolutional Neural Network (CNN) \cite{726791} that creates embeddings for each region, 2) a Self-Attention mechanism \cite{10.5555/3295222.3295349} that assigns weights between regions for functional connectivity and 3) A GNN that uses regions (nodes) and edges for graph classification. In this section we explain the purpose and details of each part separately.
Refer to Figure \ref{fig:BrainGNNArch} for the complete architecture diagram of BrainGNN.

\subsubsection{CNN Encoder}
\label{Encoder}
We use a CNN \cite{KIRANYAZ2021107398} encoder to obtain the representation of individual regions created in the preprocessing step outlined in \ref{preprocessing}. Each region vector of dimension $\tp =160$ is passed through multiple layers of one dimensional convolution, and a fully connected layer to get final embedding. The one dimensional CNN encoder used in our architecture consists of $4$ convolution layers with filter size $(4,4,3,1)$, stride $(2,1,2,1)$ and output channels $(32,64,64,10)$. This is followed by a fully connected layer resulting in a final embedding of size $64$. We use rectified linear unit (ReLU) as an activation layer between convolution layers.
Each region is encoded individually to later on create connections between regions and interpret which regions are more important/informative for classification. Our one dimensional CNN layer embeds the temporal features of regions and the spatial connections are handled in the attention and GNN parts of the architecture.

\subsubsection{Self Attention}
\label{Self-Attention}
Using the embeddings created by the CNN encoder, we estimate the connectivity between the regions of the brain using multi-head self-attention following \cite{10.5555/3295222.3295349} . The self-attention model creates three embeddings namely (key, query, value) for each region, which in our architecture are created using three simple linear layers. Each linear layer $\phi$ is of size $24$. $key_i = \phi_k(region_i)$, $query_i = \phi_q(region_i)$, $value_i = \phi_v(region_i)$.
To create weights between a region and every other region, the model takes dot product of a region's query with every other region's key embedding to get scores between them. Hence, $score_{ij} = query_i \cdot key_j$. The scores are then converted to weights using softmax.
$w_{i} = Softmax(score_i)$ where $score_i \in \R^{1 \times r}$ is a vector of scores between region $i$ and every other region.
The weights are then multiplied with the $value$ embedding of each region and summed together to create new representation for a $region_i$. Following equations show how to get new region embedding and weight values.
\begin{equation}
\begin{aligned}
&{key_i} = {region_i * W^{(k)}} , \hspace{.25cm}  {value_i} = {region_i * W^{(v)}},\hspace{.25cm}   
{query_i} = {region_i * W^{(q)}}\\
&{K} = ||_{i = 1}^r {key_i^T} = {key_i^T} || .... || {key_r^T},\hspace{.25cm}  
{weight_i} = softmax({query_i} * {K})\\
&{new\_region_i} = \sum_j^r{({weight_{ij}} * {value_j})}
\end{aligned}
\end{equation}

This process is carried out for all the regions, producing new representation of every region and the weights between regions.
These weights are then used as the functional connectivity between different regions of brain for every subject. The self attention layer encodes the spatial axis for each subject and provides with the connection between regions. The weights are learned via end to end learning of our model performing classification. This frees us from using predefined models or functions to estimate the connectivity.

\subsubsection{GNN}
\label{GNN}
Our graph network is based on a previously published model \cite{li2016gated}. Each subject is represented by a graph $G$ having ${V, A, E}$ where $V \in \R^{r \times \tp}$ is the matrix of vertices, where each vertex is represented by an embedding acquired by self-attention. $A, E \in \R^{r \times r}$ are the adjacency and edge weight matrices. Since we do not use any existing method of computing edges, we construct a complete directed graph with backward edges, meaning every pair of vertices is joined by two directed edges with weights $e_{ij}$ and $e_{ji} \in E$. For each GNN layer, at every step $s$, each node, which is a region in our model sums feature vectors of every other region relative to the weight edge between the nodes and pass the resultant and it's own feature vector through a gated recurrent unit (GRU) network \cite{cho-etal-2014-properties}, to obtain new embedding for itself.

\begin{equation}
x_s^{n_i} = \GRU (x_{s-1}^{n_i}, \sum_{\forall n_j: n_j -> n_i}e_{ji}x_{s-1}^{n_j})
\label{GNN_equation}
\end{equation}
where $\GRU$ can be explained by following set of equations, with $h_{s-1}$ representing the result of sum in Equation \ref{GNN_equation}:

\begin{equation}
\begin{aligned}
{z_s}& = \sigma (\mathbf{W}^{(z)}x_{s-1} + \mathbf{U}^{(z)}h_{s-1})\\
{r_s}& = \sigma (\mathbf{W}^{(r)}x_{s-1} + \mathbf{U}^{(r)}h_{s-1}) \\
{h_s}^{\prime}& = \sigma (\mathbf{W}x_{s-1} + r_s \odot \mathbf{U}h_{s-1})\\
{x_s}& = \sigma (z_s  \odot h_{s-1} + (1-z_s)  \odot h_{s}^{\prime})
\end{aligned}
\end{equation}

The number of steps is a hyper-parameter which we have set it as $2$ based on our experiments. The graph neural network helps nodes to create new embeddings based on the embeddings of other regions in the graph weighted by the edge weights between them. In our architecture, we use $6$ GNN layers, as shown in experiments of \cite{bresson2017residual} that it provides with the highest accuracy, with the first $3$ followed by a top-k pooling layer \cite{gao2019graph, knyazev2019understanding}. On the input feature vectors which are the embeddings of the regions, the pooling operator learns a parameter ($\mathbf{p}$) which is to assign weight to the features. Based on this parameter, top (k) layers are chosen in each pooling layer and the rest of the regions are discarded from further layers. The pooling method can be explained by the following equations.
\begin{equation}
\begin{aligned}
{y}& = \frac{{X}{p}}{\| {p} \|}\\ 
{i}&= \mathrm{top}_k({y})\\
{X}^{\prime}& = ({X} \odot  \mathrm{tanh}({y}))_{{i}},\\
{A}^{\prime}&= {A}_{{i},{i}}
\end{aligned}
\end{equation}

$\mathbf{X}^{\prime}$ and $\mathbf{A}^{\prime}$ are the new features and adjacency matrix we get after selecting top (k) regions. Pooling is performed to help model focus on the important regions/nodes which are responsible for classification. The ratio of nodes to keep in the pooling layer is a hyper-parameter and we have used $(0.8, 0.8 ,0.3)$ as the ratios.
Since we represent each subject as graph $G$, in the end we do graph classification by pooling all the feature vectors of the remaining $23$ regions/nodes.
To get one feature vector from the entire graph we concatenate the output of three different pooling layers. We pass the complete graph into three separate pooling layers. Each of the pooling layer gives us one feature factor. In the end, we concatenate the three vectors to get one final embedding for the entire graph which represents a subject. In our model we use graph max pool, graph average pool and attention based pool \cite{vinyals2016order}. The dimension of the resulting vector is $96$. The feature vector is then passed through two linear layers of size $32$ and $2$. As the name suggests, graph max pool and graph average pool just gets the max and average vector from the graph whereas attention based pooling multiplies each vector with a learned attention value before summing all the vectors.

\subsubsection{Training and Testing}
To train, validate and test our model we divide the total $311$ subjects into three groups of size $215$, $80$ and $16$, for training, validating and testing respectively. To conduct a fair experiment we use $19$ fold cross validation and for each fold we perform $10$ trials, resulting in a total of $190$ trials, and selecting $100$ subjects per class for each trial. We calculate the area under the ROC (receiver operating characteristic) curve (AUC) for each trial. To optimize our model we train all of our architecture in an end to end fashion, using Cross Entropy to calculate our loss by giving true labels $Y$ as targets, Adam as our optimizer and reducing learning rate on plateau with patience 10. We early stop our model based on validation loss, with patience of 15. Let $\theta$ represent the parameters of the entire architecture.

\begin{align}
\label{lossequation}
loss &= \CE(\hat{Y}, Y ) \\
\theta^* &= \argmin_\theta (loss; \theta)
\end{align}

\section{Results}

We show three different groups of results in our study. 1) The classification results, 2) Regions' connectivity and 3) Key regions selection. We discuss these in the following sections. We test and compare our model against the classical machine learning algorithms and ~\cite{10.1007/978-3-030-59728-3_40} on the same data used in BrainGNN. The input for the machine learning model is sFNC matrices produced using Pearson product-moment correlation coefficients (PCC).

\subsection{Classification}

\begin{figure}[H]
\centering
\includegraphics[width=0.5\linewidth]{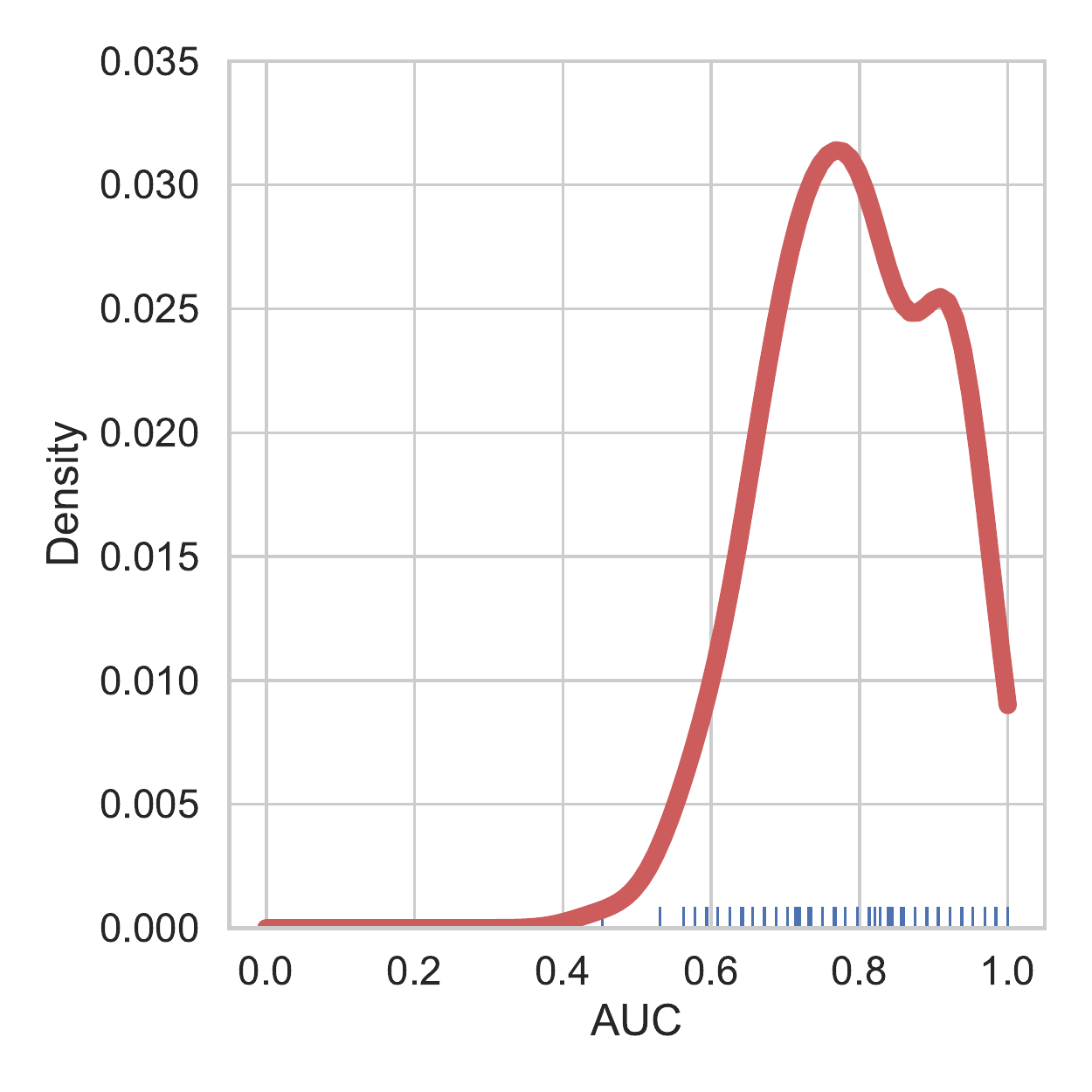}
\caption{KDE plot of probability density of ROC-AUC score on FBIRN dataset. The 190 points on the x-axis signifies the 19 fold cross validation, 10 trials per cross validation. With average and median of (($\sim 0.8$)), density peaks at ($\sim 0.8$) AUC.}
\label{fig:AUC_FBIRN}
\end{figure}

\begin{figure}[H]
\centering
\includegraphics[width=0.65\linewidth]{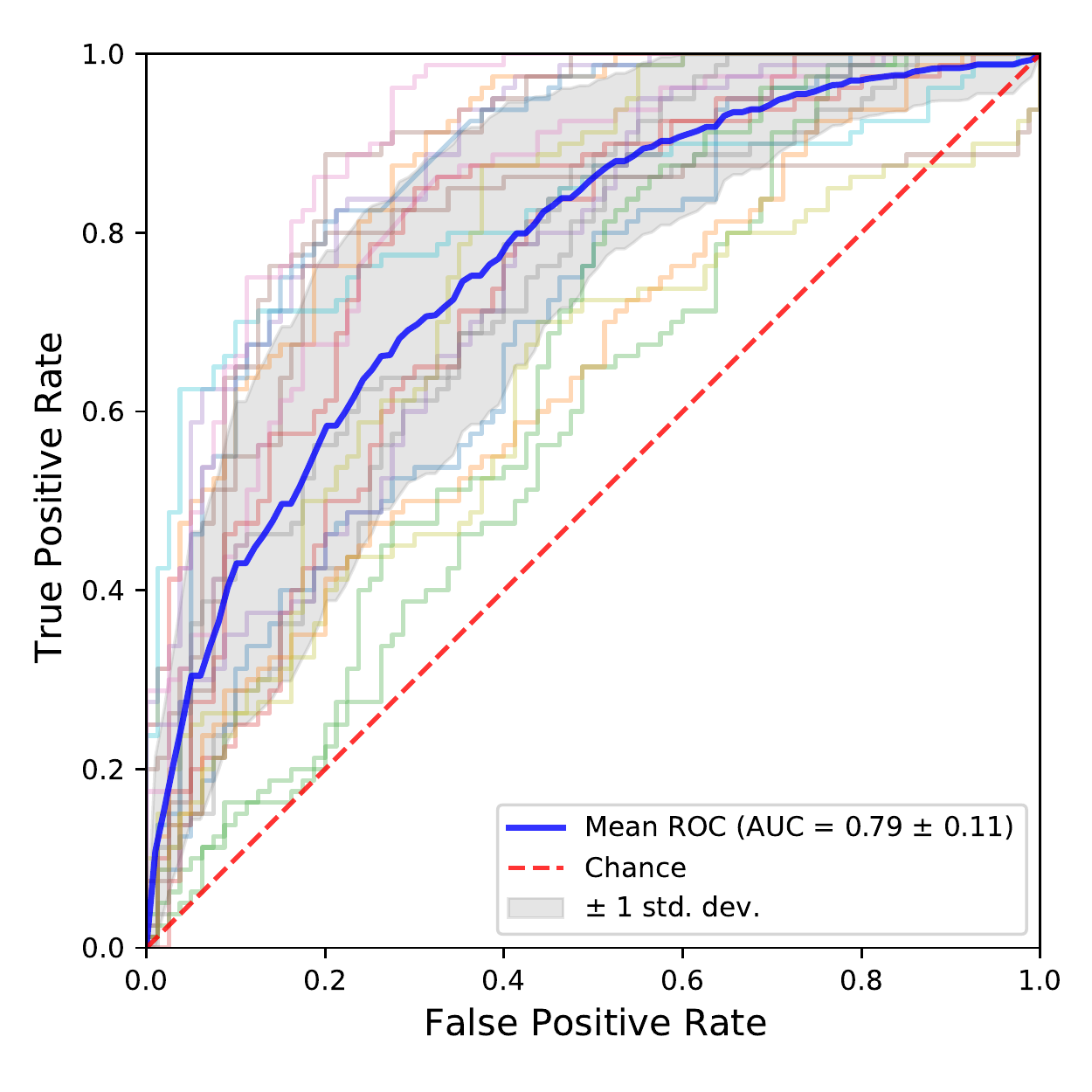}
\caption{Shows the ROC curves of the 19 models generated using each fold of cross validation. The graph is symmetrical and well balanced. It shows that the model did not learn one class over the other. }
\label{fig:ROC_FBIRN}
\end{figure}

\begin{figure}[H]
\centering
\includegraphics[width=\linewidth]{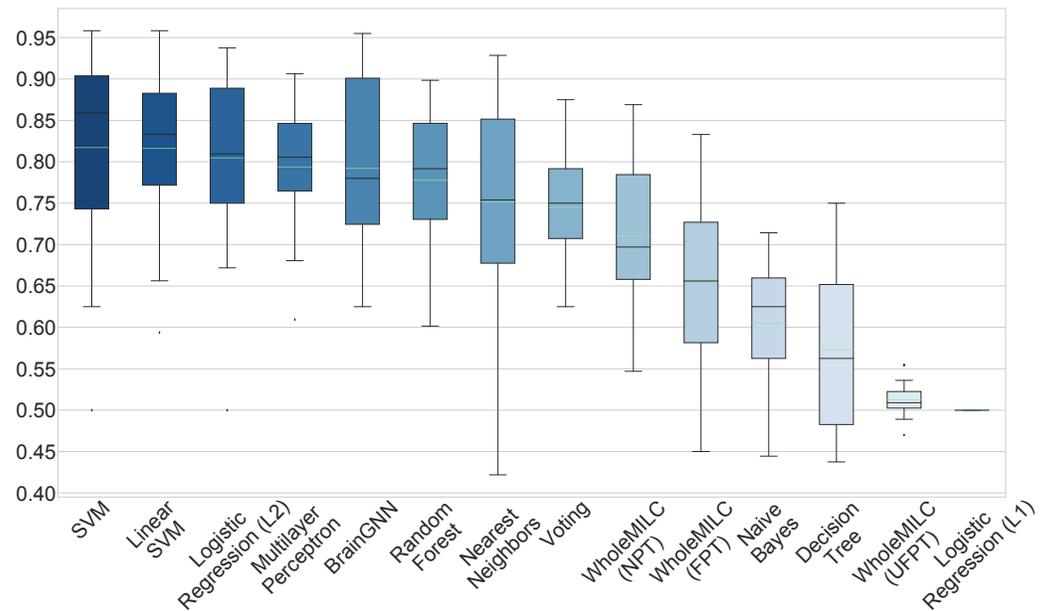}
\caption{BrainGNN comparision with other popular methods. BrainGNN provides mean AUC as $0.79$, which is just ($\sim 0.02$) less than the best performing model (SVM). Methods like WholeMILC (UFPT) and l1 logistic regression failed to learn on the input data. The l1 logistic regression model does perform better with a very weak regularization term.}
\label{fig:AUC_FBIRN_Comparision}
\end{figure}

As mentioned, we use the AUC metric to quantify the classification results of our model. AUC is more informative than simple accuracy for binary classification as in our case. Figure~\ref{fig:AUC_FBIRN} shows the results for our model. Figure~\ref{fig:ROC_FBIRN} shows the ROC curves of the models for each fold.
The performance is comparable to state of the art classical machine learning algorithms using hand crafted features and existing deep learning approaches such as ~\cite{10.1007/978-3-030-59728-3_40}, which performed test on independent component analysis (ICA) components with a hold out dataset. 
Comparison with other machine and deep learning approaches is shown in Figure~\ref{fig:AUC_FBIRN_Comparision} and prove our claim of BrainGNN providing state of the art results. BrainGNN gives almost the same mean AUC as the best performing model i.e. SVM (Support Vector Machine). To the best of our knowledge, these results are currently among the best on the unmodified FBIRN fMRI dataset \cite{RASHID2016645,Saha2020.06.24.161745,SALMAN2019101747}.  
Table~\ref{table:meanAUCtable} shows the mean AUC for each cross validation fold that was used for experimentation for BrainGNN.
As it is shown in the table that AUC has high variance across the different test sets of cross validation.
To make more sense out of the functional connectivity and region selection, both results are based on the second test fold which gives the highest ($\sim 1$) AUC score.

\subsection{Functional Connectivity}

\begin{figure*}
\centering
\begin{tabular}{c|c}
\multicolumn{1}{c|}{BrainGNN (Directed)} & \multicolumn{1}{c}{sFNC (Undirected)}
\\ \hline
\includegraphics[width=0.45\linewidth]{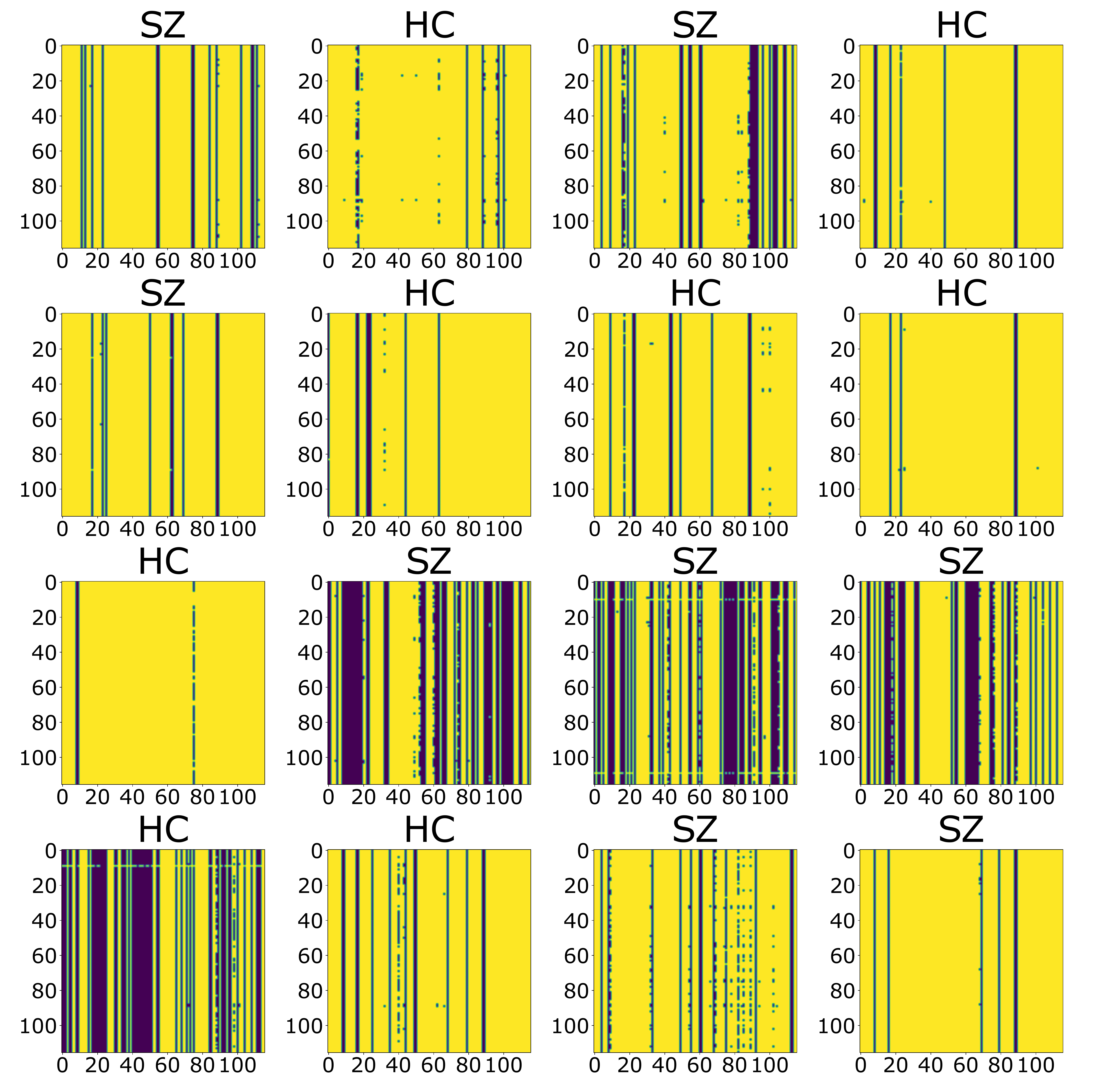} &
\includegraphics[width=0.45\linewidth]{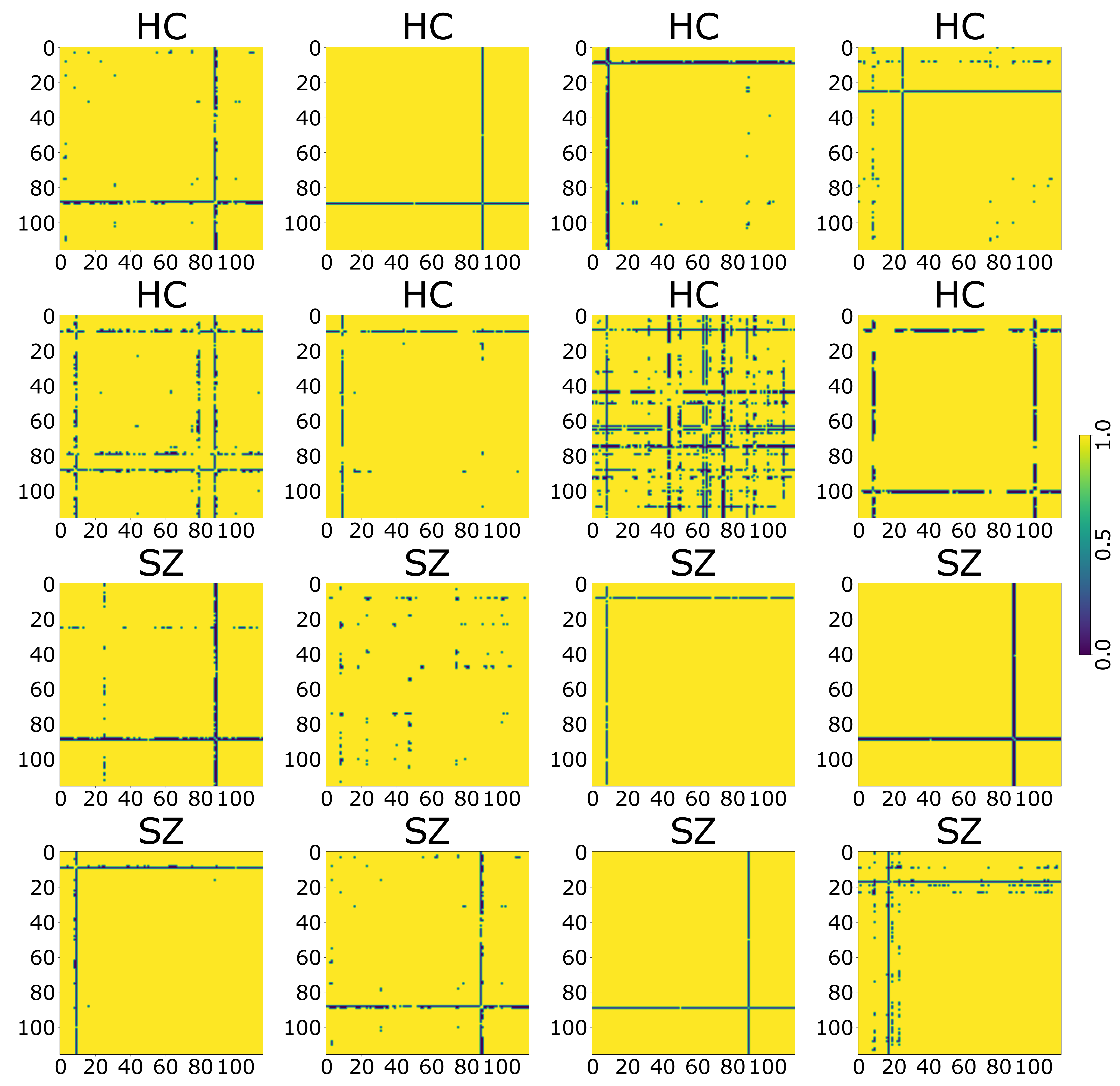} 
\end{tabular}
\caption{Connectivity between regions of subjects of both classes using BrainGNN and sFNC (PCC method). \textbf{BrainGNN: }The similarity of connection between a class and difference across class is compelling. Weights of SZ class are more sparse than HC, highlighting the fact that fewer regions receive higher weights for subjects with SZ. Refer to Table \ref{table:tTest} for results of statistical testing between weights of HC and SZ subjects. \textbf{sFNC}: The matrices are symmetric but are less informative than those produced by BrainGNN. Most of the regions are assigend unit weight.}
\label{fig:connection_weights}
\end{figure*}

The functional connectivity between regions of the brain is crucial for understanding how different parts of brain are interacting with each other. We use the weights assigned by the self-attention module of our architecture as the connection between regions. Figure \ref{fig:connection_weights} shows weight matrices for the second test set in cross validation. Weight matrices of subjects belonging to SZ class turn out to be much sparser than weights of healthy controls subjects. The result shows that the connectivity is limited to fewer regions, and functional connectivity differs across classes and fewer regions get higher weights in case of SZ subjects. We also perform statistical testing to confirm that the weight matrices of HC differ from those of SZ subjects. We create two sets each representing the concatenation of the weights of $8$ test subjects belonging to a class. We perform $2$ different testing, shown in Table \ref{table:tTest}. P-value of $<0.0001$ shows that we can reject the null-hypothesis, hence making it highly likely that the difference between weights of HC and SZ subjects is not zero.
FNC matrices produced using PCC method, do not provide such level of information and almost all regions get unit weight between other regions. \ref{fig:connection_weights} shows the usefulness of learning connectivity between regions in an end-to-end manner while training the model for classification.

\begin{table*}
\caption{Showing mean AUC of 10 trials for each cv fold}
\centering
\resizebox{\textwidth}{!}{%
\begin{tabular}{@{}llllllllllllllllllll@{}}
\toprule
\textbf {CV Fold} & 1     & 2     & 3     & 4     & 5     & 6     & 7     & 8     & 9     & 10    & 11    & 12    & 13    & 14    & 15    & 16    & 17    & 18    & 19    \\ \midrule
\textbf {AUC}     & 0.695 & 0.955 & 0.644 & 0.752 & 0.908 & 0.917 & 0.894 & 0.803 & 0.649 & 0.805 & 0.922 & 0.699 & 0.625 & 0.780 & 0.794 & 0.766 & 0.914 & 0.750 & 0.777 \\ \bottomrule
\end{tabular}%
}
\label{table:meanAUCtable}
\end{table*}

\subsection{Region Selection}
\label{regions selection}

\begin{figure}[H]
\centering
\subfloat[Region frequency]{
\includegraphics[width=0.45\linewidth]{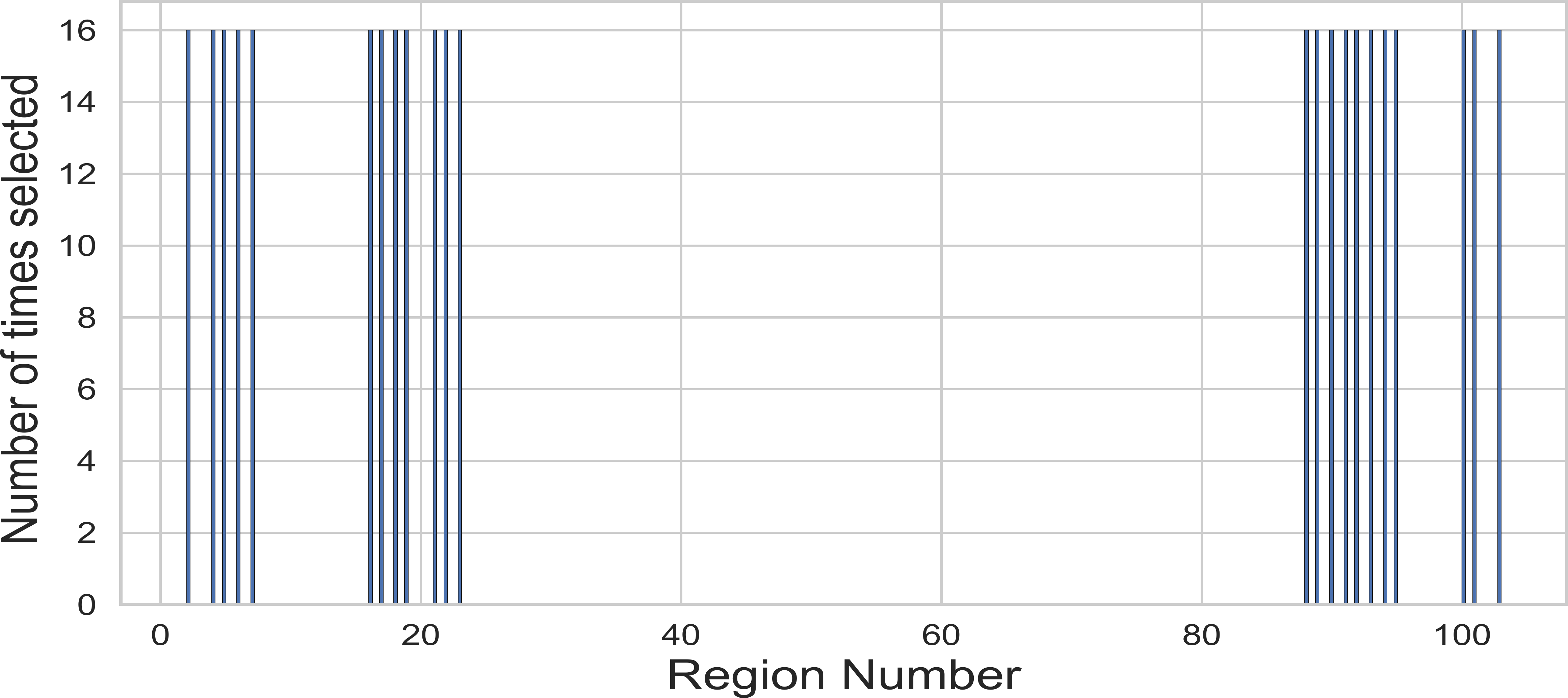}
\label{fig:rois_freq}
}
\subfloat[Region maps]{
\includegraphics[width=0.45\linewidth]{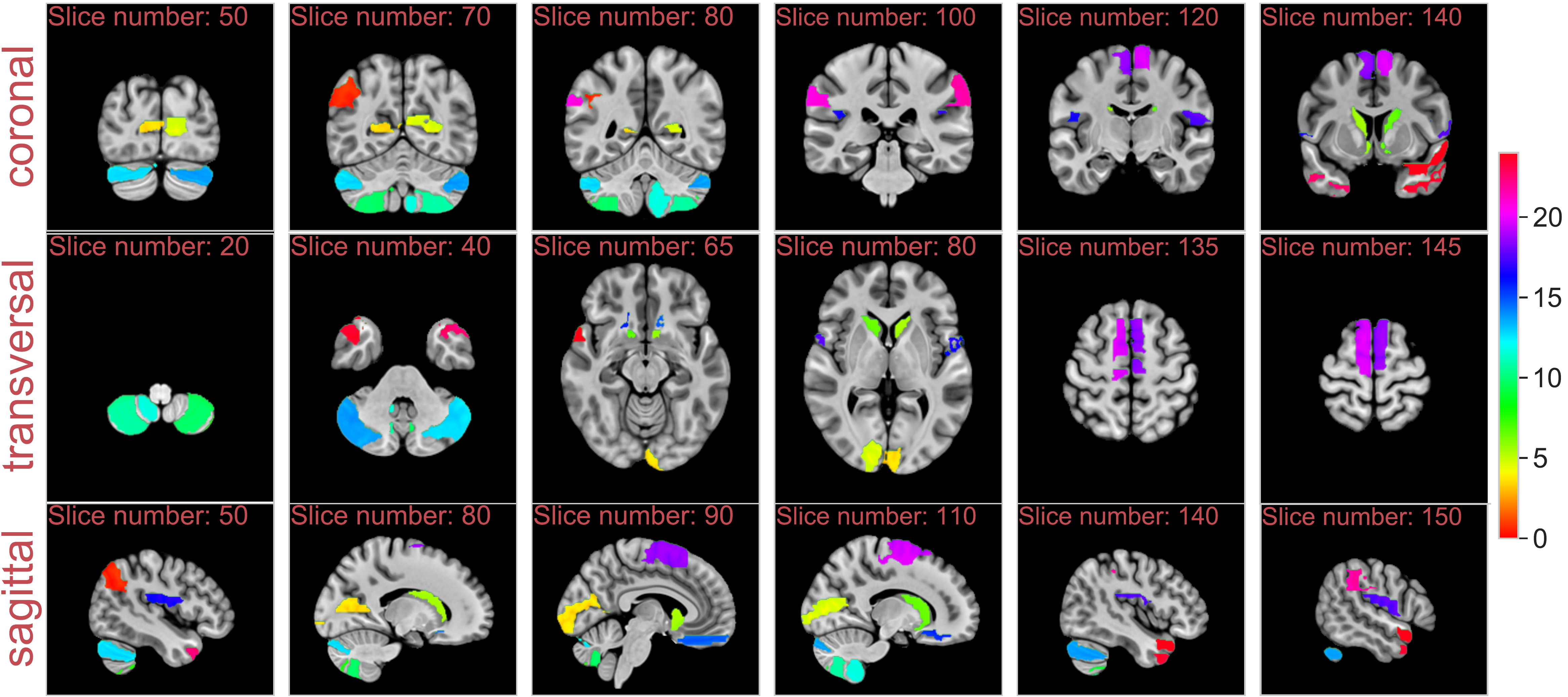}
\label{fig:maps}
}
\caption{\ref{fig:rois_freq}: Histogram of regions selected after the last pooling layer of GNN. $2^{nd}$ fold of the cross validation gives this figure. All $23$ regions are selection equal number of times ($16$). It further signifies the important of these regions, showing that for all subjects across both classes, these $23$ regions are always selection. \ref{fig:maps}: Mapping the $23$ regions back on the brain across the three anatomical planes. $100^{th}$ time point is selected for these brain scans. X axis shows different slices of the plane.}
\label{fig:regions_histogram}
\end{figure}

The pooling layer added in our GNN module allows us to reduce the number of regions. Functionality across brain regions differ significantly and not all regions are affected by a disorder or have any noticeable affect on classification. This makes it very important to know which regions are more significantly informative of the underlying disorder and study how they get affected or affect the disorder. Figure \ref{fig:rois_freq} shows the final $23$ regions selected after the last pooling layer in the GNN model which is just $20$ percent of the total brain regions used. The relevance of these regions is further signified by the fact that the graph model has no residual connections and the final feature vector created after the last GNN layer is through the feature vectors of these regions. Figure \ref{fig:maps} shows the location of the selected regions in the MNI brain space, regions are distinguished by color. Each region is assigned one unit from the color bar, used to represent signal variation in the fMRI data.

\begin{table*}[]
\centering
\caption{Statistical testing between weight matrices of HC and SZ. The test shows that weights of regions differ across HC and SZ subjects. Refer to Figure~\ref{fig:AUC_FBIRN_Comparision} for mean and deviation of these folds.}
\begin{tabular}{ll}
\hline
Test & P Value \\ \hline
Mann-Whitney U Test & 0.0 \\ \hline
Welch's t-test & 0.0
\end{tabular}
\label{table:tTest}
\end{table*}

\section{Discussion}

The richness of results in the three presented categories highlights the benefits of the proposed method.
High classification performance shows that the model can accurately classify the subjects and hence can be trusted with the other two interpretative results of the paper. Functional connectivity between regions shown in the paper is of paramount importance as it highlights how brain regions are connected to each other and the variation between classes. Learning functional connectivity end-to-end through classification training frees the model from depending on an external method. The sparse weight matrix of subjects with SZ shows that connectivity remains significant between considerably fewer regions than for healthy controls.
Notably, the attention based functional connectivity cannot be interpreted as the conventional correlation based symmetric connectivity.
  Due to the inherent asymmetry in keys and values the obtained graph is directed but is also prediction based rather than simply correlation.
  We expect that a further investigation into the obtained graph structure will bring more results and deeper interpretations. The sparsity is to be further explored and seen in context of the regions selected, shown in the last section of results.
The final regions selected by the model strengthens our hypotheses that not all regions are equally important for identifying a particular brain disorder.
Reducing the brain regions by almost $80\%$ helps in identifying the important regions for classification of SZ.
The regions selected by our model such as (cerebellum, temporal lobe, caudate, SMA) etc have been linked to the disease by multiple previous studies, hence reassuring the correctness of our model \cite{article, https://doi.org/10.1002/hbm.25205, article2, article3}.
We see an immediate benefit of using GNNs to study functional connectivity and our BrainGNN model specifically. The data-driven model almost eliminates manual decisions transitioning graph construction and region selection into the data-driven realm. With this BrainGNN opens up a new direction to the existing studies of connectivity and we expect further model introspection to yield insight into the spatio-temporal biomarkers of schizophrenia.
Further reducing the selected regions and how they different across subjects belonging to different class is also left for future work.
We envision great benefits to interpretability and elimination of manual processing and decisions in a future extension of the model that would enable it to work directly from the voxel-level not only connecting and selecting ROIs, but also constructing them.





\vspace{6pt}



\authorcontributions{``Conceptualization, U.M., S.P.; methodology, U.M.; software, U.M.; validation, U.M.; formal analysis, U.M.; investigation, U.M.; resources, V.C.; data curation, Z.F, U.M.; writing--original draft preparation, U.M.; writing--review and editing, U.M., S.P., Z.F.; visualization, U.M.; supervision, S.P.; project administration, S.P., V.C.; funding acquisition, S.P., V.C. All authors have read and agreed to the published version of the manuscript.'', please turn to the \href{http://img.mdpi.org/data/contributor-role-instruction.pdf}{CRediT taxonomy} for the term explanation. }

\funding{This study was in part supported by NIH grants 2RF1MH121885, 1R01AG063153, and 2R01EB006841.}

\acknowledgments{Data for Schizophrenia classification was used in this study were downloaded from the Function
BIRN Data Repository (http://bdr.birncommunity.org:8080/BDR/),
supported by grants to the Function BIRN (U24-RR021992) Testbed funded by the
National Center for Research Resources at the National Institutes of Health, U.S.A. and from the COllaborative Informatics and Neuroimaging
Suite Data Exchange tool (COINS; \href{http://coins.trendscenter.org}{http://coins.trendscenter.org}) and data collection was performed at the
Mind Research Network, and funded by a Center of Biomedical Research Excellence (COBRE)
grant 5P20RR021938/P20GM103472 from the NIH to Dr.Vince Calhoun.}

\conflictsofinterest{The authors declare no conflict of interest.}







\end{paracol}
\reftitle{References}



\externalbibliography{yes}
\bibliography{GNN}






\end{document}